\documentclass[preprint,12pt]{elsarticle}

\usepackage[cmex10]{amsmath}
\usepackage{amssymb}
\usepackage{amsthm}
\usepackage{overpic}
\usepackage{url}

\usepackage{lineno}

\def\imagunit{\mathsf{j}} 
\newcommand{\vect}[1]{\boldsymbol{#1}}

\journal{IEEE Signal Processing Cup 2021}

\begin{document}

\begin{frontmatter}

\title{\large{IEEE Signal Processing Cup 2021}\\[5mm] \Large{Configuring an Intelligent Reflecting Surface for Wireless Communications}}

\author{Emil Björnson}

\address{KTH Royal Institute of Technology, Sweden \\ Linköping University, Sweden\\emilbjo@kth.se\\[5mm] Official website: \url{https://github.com/emilbjornson/SP_Cup_2021}}

\begin{abstract}
The IEEE Signal Processing Society is proud to announce the eighth edition of the Signal Processing Cup: an exciting challenge to control a wireless propagation environment using an intelligent reflecting surface. 

An intelligent reflecting surface is a two-dimensional array of metamaterial whose interaction with electromagnetic waves can be controlled, e.g., by tuning the impedance variations over the surface. These surfaces might be used in the sixth generation (6G) mobile technology to direct wireless signals from a transmitter towards a receiver, to raise the communication performance. The goal of the challenge is to characterize the behavior of an intelligent reflecting surface based on received signals from an over-the-air signaling phase and develop a control algorithm to configure the surface to aid wireless communications.
\end{abstract}

\end{frontmatter}

\section{Introduction}
\label{S:1}

\begin{figure*}[t!]
	\centering 
	\begin{overpic}[width=\columnwidth,tics=10]{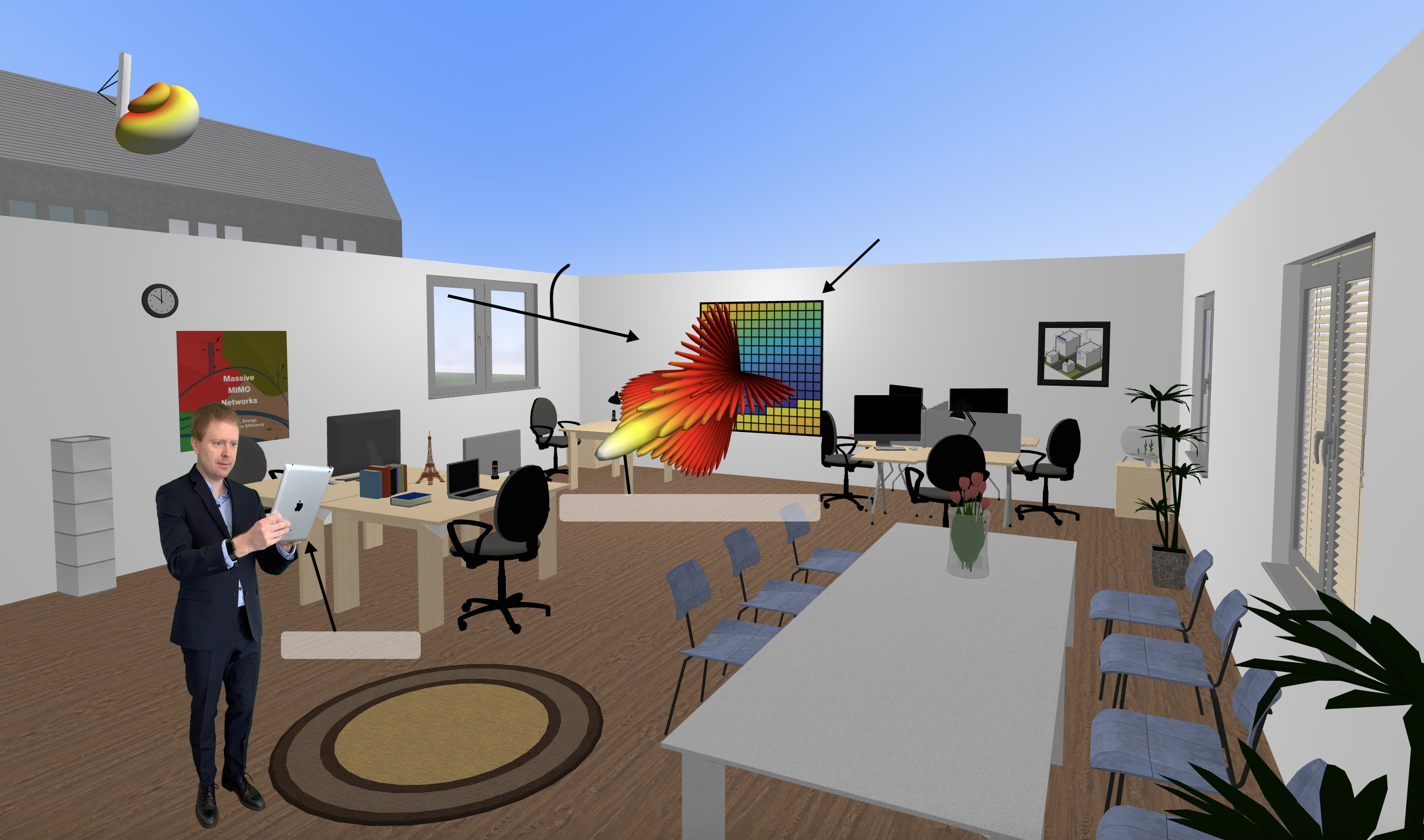}
	\put(8,56.5){\scriptsize 1) Transmitting base station}
	\put(37,41){\scriptsize 2) Signal reaching IRS}
	\put(60,47){\scriptsize 3) IRS with elements}
	\put(60,45){\scriptsize configured to shape a beam}
	\put(60,43){\scriptsize towards the receiver}
	\put(39.5,22.8){\scriptsize 4) Beam from IRS}
	\put(20,13.2){\scriptsize 5) User}
\end{overpic} 
	\caption{A typical use case of an IRS, where it receives a signal from a transmitting base station and re-radiates it focused towards the receiving user. To focus the beam in the right direction, the IRS must be configured properly.}
	\label{figureBasicExample} 
\end{figure*}
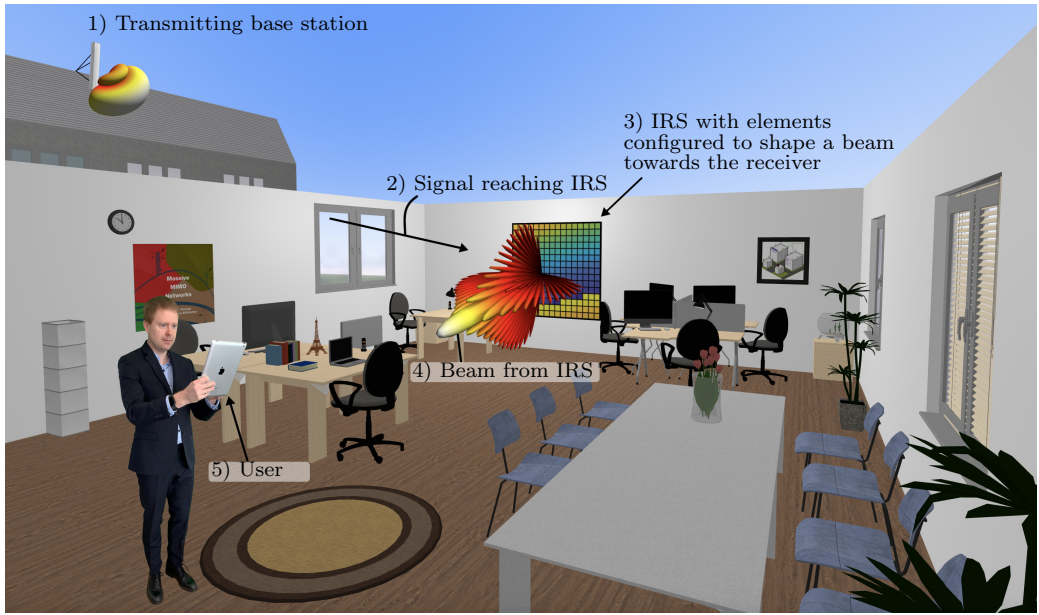

\noindent This competition considers a communication setup where a single-antenna base station has been deployed to serve user devices that are located in a particular coverage area. A high-level illustration of this use case is provided in Figure~\ref{figureBasicExample}, where there are plenty of objects in between the base station and the area where the prospective users reside. The radio waves interact with these objects in a way that traditionally is uncontrollable. In the figure, the direct path between the base station and the user is blocked by a wall and thus highly attenuated, leading to a low signal-to-noise ratio (SNR). To improve the propagation conditions, an intelligent reflecting surface (IRS) has been deployed at a fixed location on a wall. The IRS consists of $N=4096$ elements (also known as meta-atoms or unit cells in the literature), which are passive in the sense that they will (diffusely) reflect radio waves that are reaching them from the base station, but will not process or amplify the signals. However, each IRS element has a tunable impedance which is controlled using a switch that has two states in the considered setup. By changing the switch, the reflection coefficient can be changed, which implies that the amplitude and phase-shift of the reflected signal are modified. Apart from the ability to modify the impedance, the IRS is passive, which implies that there is no added noise or distortion.

The general principle of IRS-aided communication is to set the switches in such a way that the signals from the $N$ elements are constructively combined at the user location. In other words, the diffusely reflected signals are phase-shifted so that constructive interference occurs in the direction of the receiver, which results in the beam shown in Figure~\ref{figureBasicExample}.
The main challenge is to select which configuration to utilize for each given user. The input to such a selection algorithm is measurements of the received signals at the user for different configurations. The IRS cannot make any measurements on its own (since it is passive), but can only receive a control signal (from the base station or user) that tells it which configuration to use.
There are $2^{N} = 2^{4096}$ possible configurations in the considered setup which is, practically speaking, infinitely large. Hence, an exhaustive search is hopeless, both from the measurement and computational perspectives.

The goal of this competition is to design an algorithm that selects a good IRS configuration based on measurements made at the user location. It is plausible to design such algorithms based on a number of measurements that grows linearly with $N$, rather than exponentially, since the number of unknown channel parameter is proportional to $N$. To show this, we need to develop a system model for the considered setup.

\section{System model}

\noindent  The transmission over the communication channel is carried out using orthogonal frequency-division multiplexing (OFDM). The signal is generated using pulse-amplitude modulation (PAM) with a unit-energy sinc-function as the pulse-shape filter and the received signal is matched filtered and sampled at the symbol rate. 

Let $\{ x[k] \}$ denote the transmitted discrete-time signal in the complex baseband domain. The corresponding received discrete-time signal is $\{ z[k] \}$ in the complex baseband and is given by
\begin{equation}
z[k] =  \sum_{\ell=0}^{M-1} h_{\vect{\theta}}[\ell] x[k-\ell] + w[k]
\end{equation}
where $\{ h_{\vect{\theta}}[\ell] : \ell=0,\ldots,M-1 \}$ is the finite impulse response (FIR) filter that describes the communication channel and  $\{ w[k] \}$ is the receiver noise. The FIR filter has $M$ coefficients, which depend on the physical channels as well as the IRS configuration, where the latter is denoted by $\vect{\theta}$.

The OFDM transmission makes use of an $M-1$ length cyclic prefix and creates $K>M$ subcarriers. Hence, a block of $K+M-1$ time-domain signals are transmitted to create one OFDM block with $K$ parallel subcarriers:
\begin{equation} \label{eq:system-model}
\bar{z}[\nu] = \bar{h}_{\vect{\theta}}[\nu] \bar{x}[\nu] + \bar{w}[\nu], \quad \nu = 0, \ldots, K-1
\end{equation}
where
\begin{align} 
\bar{z}[\nu] &= \frac{1}{\sqrt{K}} \sum_{k=0}^{K-1} z[k] e^{-\imagunit 2 \pi k \nu /K}, \\
\bar{x}[\nu] &= \frac{1}{\sqrt{K}} \sum_{k=0}^{K-1} x[k] e^{-\imagunit 2 \pi k \nu /K} ,\\
\bar{h}_{\vect{\theta}}[\nu] &= \sum_{k=0}^{M-1} h_{\vect{\theta}}[k] e^{-\imagunit 2 \pi k \nu /K}, \\
\bar{w}[\nu] &= \frac{1}{\sqrt{K}} \sum_{k=0}^{K-1} w[k] e^{-\imagunit 2 \pi k \nu /K}.
\end{align}
If we let $\odot$ denote the Hadamard (element-wise) product, we can write the system model in \eqref{eq:system-model} in vector form as:
\begin{equation}
\underbrace{\begin{bmatrix} \label{eq:system-model2}
\bar{z}[0]  \\
\vdots \\
\bar{z}[K-1] 
\end{bmatrix}}_{=\vect{\bar{z}}} = \underbrace{\begin{bmatrix}
\bar{h}_{\vect{\theta}}[0]  \\
\vdots \\
\bar{h}_{\vect{\theta}}[K-1] 
\end{bmatrix}}_{=\vect{\bar{h}}_{\vect{\theta}}} \odot \underbrace{\begin{bmatrix}
\bar{x}[0]  \\
\vdots \\
\bar{x}[K-1] 
\end{bmatrix}}_{=\vect{\bar{x}}} + \underbrace{\begin{bmatrix}
\bar{w}[0]  \\
\vdots \\
\bar{w}[K-1] 
\end{bmatrix}}_{=\vect{\bar{w}}}.
\end{equation}
This is the system model of one OFDM block and the IRS configuration is constant within the block. The input-output relation can be written in short form as $\vect{\bar{z}} = \vect{\bar{h}}_{\vect{\theta}} \odot \vect{\bar{x}} + \vect{\bar{w}}$. Note that all the vectors are of length $K$.

Different configurations $\vect{\theta}$ result in different channel vectors $\vect{\bar{h}}_{\vect{\theta}}$ and, eventually, different communication performance. To select a good IRS configuration, we can try different configurations and estimate the corresponding channels.
To estimate the channel and IRS properties that are determining the unknown vector $\vect{\bar{h}}_{\vect{\theta}}$, one can transmit a known vector $\vect{\bar{x}}$, often referred to as a pilot signal, and try to extract $\vect{\bar{h}}_{\vect{\theta}}$ from the noisy received signal $\vect{\bar{z}}$.

Since the channel vector $\vect{\bar{h}}_{\vect{\theta}}$ contains $K$ entries and there are $K$ received signals in \eqref{eq:system-model2}, there are sufficient degrees-of-freedom to estimate all of these signals. However, the vector $\vect{\bar{h}}_{\vect{\theta}}$  depends on both the propagation channel and the IRS configuration $\vect{\theta}$. To determine a suitable configuration, multiple received signals obtained using different IRS configurations must be obtained.
Assuming that the propagation channel is constant, since there are $N$ elements, it should be sufficient to consider a sequence of $N$ configuration $\vect{\theta}_1,\ldots,\vect{\theta}_N$ that are carefully selected.

The task of this Signal Processing Cup is to develop an algorithm that selects a suitable IRS configuration for a given dataset of received signals obtained as described above. A dataset with the received signals for different users is provided. Since the described procedure only explores $N$ out of $2^N$ configurations, it is likely that the best choice is not among those that were actually tried out during the pilot transmission.

When the IRS has been properly configured, the vector $\vect{\bar{x}}$ can contain unknown data that is transmitted to the receiving user. The overall goal of an IRS-aided system is to obtain a high communication performance after the IRS has been configured. Hence, the winning team of this Signal Processing Cup will be the one that provides the highest communication performance.

\subsection{Suggested references} \label{subsec:references}

To learn more about the basic system modeling, we recommend the following references:

\begin{itemize}

\item [Ref1] Tutorial article ``\emph{Reconfigurable Intelligent Surfaces: A Signal Processing Perspective With Wireless Applications}'' by Emil Björnson et al., \url{https://arxiv.org/abs/2102.00742}

\item [Ref2]  YouTube video: ``\emph{Reconfigurable Intelligent Surfaces: A Signal Processing Perspective}'', \url{https://youtu.be/bcSDRXoVQvk}

\item [Ref3]  Wireless Future Podcast: ``\emph{Ep 3. Reconfigurable Intelligent Surfaces}'', \url{https://youtu.be/4yfQzEGiVlY}

\end{itemize}

Note that intelligent reflecting surfaces are also being referred to  as reconfigurable intelligent surfaces and software-controlled metasurfaces.

\subsection{Evaluation criterion}

\noindent The evaluation is based on the 50 users that are considered in the dataset \texttt{dataset2.mat}, described in detail in Section~\ref{sec:dataset}. The team should identify one ``optimized'' IRS configuration per user and then submit their 50 configuration vectors $\vect{\theta}$ before the deadline. The vectors will most likely be different for each user since the surface should reflect the signal to a different location.
Based on the submitted vectors, the organizer will compute the data rates that each user achieves with those configurations. The rate is computed as
\begin{align} \label{eq:rate-OFDM}
    R = \frac{B}{K+M-1} \sum_{\nu=0}^{K-1} \log_2 \left( 1 + \frac{P | \bar{h}_{\vect{\theta}}[\nu] |^2}{B N_0} \right)
\end{align} 
where $P=1$\,W is the signal power, $B=10$\,MHz is the bandwidth (symbol rate), and $N_0$ is the noise power spectral density (which has purposely not been stated in this document, but is the same for all users).
A weighted average rate over all the users will be computed, where the weights double the rates of the non-line-of-sight users, since these typically have lower rates than the line-of-sight users. The dataset doesn't declare which users belong to which category.

The teams that hand in understandable high-level descriptions of their algorithms will be ranked based on the highest weighted average rate.

\vspace{-3mm}

\section{Details about the setup and dataset} \label{sec:dataset}

\noindent Since this is an emerging technology with limited experimental equipment, the data has been generated synthetically in MATLAB. The transmitting base station and the IRS are at fixed locations and have a line-of-sight channel between them. Although the direct path is constant, there can be minor fluctuations in the scattered paths between them, due to the movements of other objects.
The users are located at different locations, but in the same coverage area, thus there might be common scattering clusters that affect multiple users. All users have non-line-of-sight channels to the base station, which motivates the use of the IRS.
Some users will have line-of-sight channels from the IRS, other users will not.
We will not disclose the exact details of how the data was generated since one of the goals of this challenge is to discover its structure. The tutorial article [Ref1] gives a good introduction to the system modeling, but the simulation setup contains further hardware-related features that are not explicitly modeled in that article.

There are $K = 500$ subcarriers and the time-domain FIR filter has $M = 20$ taps. The carrier frequency is 4 GHz, the bandwidth is $B=10$ MHz and since an ideal sinc-function is used for pulse shape, the symbol rate is $10^7$ symbols per second.

Each element of the IRS can be configured to be in one of two states: $+1$ and $-1$. You can think of them as representing ON and OFF. Recall that the vector $\vect{\theta}$ was used to represent the different states, which implies that $\vect{\theta} \in \{-1,+1\}^N$ and the $n$th entry represents the state of the $n$th IRS element.
The states will result in different impedances in the element, which in turn results in different reflection coefficients. The hardware designer claims that the phase-shift difference between the two states is equal to $\pi$ and that the same components have been used for all elements.

During the pilot transmission, all the elements of $\vect{\bar{x}}$ are equal. The signal is provided in the dataset. The IRS is switching between different configurations, in order to enable channel estimation and configuration. The configuration sequence $\vect{\theta}_1,\ldots,\vect{\theta}_N$ consists of the columns of an $N \times N$ Hadamard matrix, which is provided in the dataset.

\subsection{Available measurement data}

\noindent  The following datasets are provided at\\ \url{https://github.com/emilbjornson/SP_Cup_2021}

\begin{enumerate}
\item \textbf{dataset1.mat} The first dataset focuses on a single user and contains $4N$ received OFDM signal blocks, each obtained when transmitting a known pilot signal over all subcarriers using varying IRS configurations. Since there are more than $N$ received signals, this data can potentially be utilized to both learn channel-related and IRS-related channel properties. The following variables are provided:

\begin{itemize}

\item \texttt{K}: Number of subcarriers, $K=500$.
\item \texttt{M}: Number of channel taps, $M=20$.
\item \texttt{N}: Number of IRS elements, $N=4096$.
\item \texttt{pilotMatrix4N}: Matrix of size $N \times 4N = 4096 \times 16384$. Each column contains one configuration $\vect{\theta} \in \{-1,+1\}^N$ that was used during the transmission.
\item \texttt{receivedSignal4N}: Matrix of size $K \times 4N = 500 \times 16384$. Each column contains the received signal $\vect{\bar{z}}$ when transmitting the pilot signal $\vect{\bar{x}}$ using one of the configurations in \texttt{pilotMatrix4N} (the one with the matching column index).
\item \texttt{transmitSignal}: Matrix of size $K \times 1 = 500 \times 1$. The transmitted pilot signal $\vect{\bar{x}}$, which is used for each transmission.

\end{itemize}

\item \textbf{dataset2.mat} The second dataset considers 50 other users and contains $N$ received OFDM signal blocks, each obtained when transmitting a known pilot signal over all subcarriers, and using different IRS configurations. It is the communication performance of these 50 users that are considered when evaluating the communication performance according to \eqref{eq:rate-OFDM}. The following variables are provided:

\begin{itemize}

\item \texttt{K}: Number of subcarriers, $K=500$.
\item \texttt{M}: Number of channel taps, $M=20$.
\item \texttt{N}: Number of IRS elements, $N=4096$.
\item \texttt{pilotMatrix}: Matrix of size $N \times N = 4096 \times 4096$. Each column contains one configuration $\vect{\theta} \in \{-1,+1\}^N$ that was used during the transmission.
\item \texttt{receivedSignal}: Array of size $K \times N \times 50 = 500 \times 4096 \times 50$. The third dimension represents each of the 50 users. For a given user $i$, each column of \texttt{receivedSignal}$(:,:,i)$ contains the received signal $\vect{\bar{z}}$ when transmitting the pilot signal $\vect{\bar{x}}$ using one of the configurations in \texttt{pilotMatrix} (the one with the matching column index).
\item \texttt{transmitSignal}: Matrix of size $K \times 1 = 500 \times 1$. The transmitted pilot signal $\vect{\bar{x}}$, which is used for each transmission.

\end{itemize}

\end{enumerate}

\subsection{MATLAB Software}

\noindent The datasets are delivered in the \texttt{.mat} format, which can be easily imported into MATLAB. Participating students are encouraged to download the complimentary Mathworks Student Competitions Software for use in the competition: \\

\noindent {\scriptsize \url{https://mathworks.com/academia/student-competitions/software-request-registration-sp-cup.html}}

\subsection{Submission of solution}

\noindent Each team should submit its solution in the form of a \texttt{.mat} file containing a variable \texttt{theta} of dimensions $4096 \times 50$. Each entry must be either $+1$ or $-1$. Each column of this matrix is the selected IRS configuration for one of the users in \texttt{dataset2.mat}: column $l$ is the configuration for user $l$, for $l=1,\ldots,50$.

The team must also submit a text document containing the team name, the names and email addresses of all the team members, and a high-level description of the algorithm that was used to identify the solution. No simulation code or pseudo-code is required at this point. The report should be focused on describing the intuition behind the proposed solution, to give the organizer a sense of what the team has done. The description should not be longer than three pages.

The organizer will then compute \eqref{eq:rate-OFDM} using $P=1$\,W and $B=10$\,MHz. Note that the teams cannot compute the exact value of  \eqref{eq:rate-OFDM} since it depends on the true channels, which users have line-of-sight conditions, and the undisclosed value of the noise power spectral density.

\section{General details and rules about the competition}

\noindent The Signal Processing Cup (SP Cup) competition is held annually and encourages teams of students to work together to solve real-world problems using signal processing methods and techniques.

\begin{itemize}

\item \textbf{Team composition:} The purpose of this competition is that teams of 3-10 undergraduate students (enrolled as Bachelor or Master students) compete and develop a solution, under the supervision of a faculty member (or someone else with a PhD degree) and (optionally) the tutorship of a PhD student or postdoc. It is the undergraduate students that are responsible for presenting their work on ICASSP 2021, if the team makes it to the final competition (see below).

\item \textbf{Detailed eligibility requirements:}
Each team should contain 3-10 undergraduate students, who carry out the majority of the work. An undergraduate student is a person without a Master degree (or equivalent) at the time of submission. The students can be enrolled to a Bachelor or Master program, or equivalent.
At least three of the undergraduate team members must hold either regular or student memberships of the IEEE Signal Processing Society at the time of submission.  It is not mandatory to be enrolled at a university at the period of the cup.
Undergraduate students who are in the first two years of their college studies as well as high-school students who are capable to contribute are welcome to participate in a team. An undergraduate student can only be the member of one team.
The students are to be supervised by a faculty member or someone else with a PhD degree that is employed by the university. The supervisor can be assisted by a tutor who has earned at least a Master degree at the time of submission; for example, a PhD student, postdoc, or equivalent. Each person can only be a member of one team.

\item \textbf{Prerequisites:} There are no formal prerequisites but we recommend that the team members have studied linear algebra and a first course in signals-and-systems (or a similar elementary signal processing). As a first step, we recommend the team to read the first 24 pages of the tutorial article mentioned in Section~\ref{subsec:references} (you can skip ``Reconfiguration under mobility''), which is closely related to the considered problem.

\item \textbf{Dataset:} A novel dataset with transmitted and received signals, for different configurations, is provided for the challenge. See Section~\ref{sec:dataset}.

\item \textbf{Prize:} The three teams with the highest performance in the open competition will be selected as finalists and will be invited to participate in the final competition at ICASSP 2021. It was announced on January 26 that it will be a fully virtual conference, which implies that the grand final will also be organized virtually during the conference days. The champion team will receive a grand prize of \$5,000. The first and the second runner-up will receive a prize of \$2,500 and \$1,500, respectively. These prizes are sponsored by MathWorks, Inc. We are grateful for their continued support of the IEEE Signal Processing Cup.

\end{itemize}

\subsection{Important Dates}

\begin{itemize}
\item Release of the dataset and detailed instructions: February 1, 2021
\item Registration of team (see Section~\ref{subsec:registration} below): March 1, 2021
\item Submission deadline: April 26, 2021 (preliminary version) and May 9, 2021 (final version)
\item Finalists Announcement: May 11, 2021
\item Finale at ICASSP: June 6, 2021
\end{itemize}

\subsection{Registration}
\label{subsec:registration}

\noindent The team registration is done in the following IEEE system:

\url{https://www2.securecms.com/SPCup/SPCRegistration.asp}\\

At least one member per team should also sign up to the Q/A platform Piazza (see below) to get announcements about events, changes in the instructions, and summarizes of Q/A.

\subsection{Questions}

\noindent If you have any questions about the problem setup and data, please contact the organizer. The preferred way is to make use of the Q/A platform Piazza: \url{https://piazza.com/ieee_sps/spring2021/spcup2021/home}

This document will be updated if any changes in the instructions are required. It is important for all teams that enter the competition register by sending an email to the organizer, so that changes and updates can be broadcasted to all teams.

\end{document}